\RequirePackage{etex}

\documentclass[pdftex,colorlinks,notitlepage,a4paper,12pt]{article}

\usepackage[a4paper]{geometry}

\usepackage[pdftex]{graphicx}

\usepackage{epstopdf}

\usepackage[usenames,dvipsnames]{color}
\usepackage[dvipsnames,table]{xcolor}

\usepackage{soul}
\soulregister\textcite7
\soulregister\parencite7
\soulregister\ref7
\soulregister\parentext7
\soulregister\footnote7

\usepackage{pdftricks2}
\usepackage[utf8]{inputenc}
\usepackage[T1]{fontenc}

\usepackage[french, german,spanish, british]{babel}
\babeltags{es = spanish}
\babeltags{de = german}
\usepackage[style=american]{csquotes}

\usepackage[en-GB]{datetime2}
\DTMlangsetup[en-GB]{ord=raise}

\usepackage{latexsym}

\usepackage{amsmath}

\usepackage{econometrics}

\usepackage{amssymb}

\usepackage{pdflscape} % \begin{landscape} ... \end{landscape}

\usepackage{afterpage}

\usepackage{array}
\usepackage{delarray}
\usepackage{cases}
\usepackage{rotating}
\usepackage{longtable}
\usepackage{threeparttablex}
\usepackage{booktabs}
\usepackage{caption}
\usepackage{makecell}
\usepackage{tabularx}
\usepackage{ltablex}
\keepXColumns
\usepackage{tabu}
\usepackage{linegoal}

\makeatletter
\newlength\LongtableWidth
\newcommand*{\org@longtable}{}
\let\org@longtable\longtable
\def\longtable{%
	\begingroup
	\advance\c@LT@tables\@ne
	\edef\x{LT@\romannumeral\c@LT@tables}
	\global\LongtableWidth\z@
	\@ifundefined{\x}{%
	}{%
		\def\LT@entry##1##2{%
			\global\advance\LongtableWidth##2\relax
		}%
		\@nameuse{\x}%
	}%
	\typeout{* \x: \the\LongtableWidth}%
	\endgroup \ifdim\LongtableWidth>\z@ \setlength{\LTcapwidth}{\LongtableWidth}%
	\fi
	\org@longtable }
\makeatother

\usepackage{ragged2e}

\usepackage{xpatch}

\usepackage{float}
\usepackage{placeins}

\usepackage{eurosym}    % \euro{} and \euro{1}

\usepackage[hyphens]{url}

\usepackage{appendix}

\usepackage[super]{nth}

\usepackage[shortlabels]{enumitem} % a package for nice enumerations
\setlist[enumerate]{nosep, label = {(\arabic*)}}
\setlist[itemize]{nosep, label = {---}}

\usepackage{titlesec}
\titlelabel{\thetitle.\quad}
\titleformat*{\section}{\normalfont\large\bfseries\setstretch{1.5}}
\titleformat*{\subsection}{\normalfont\large\bfseries\setstretch{1}}

\usepackage{fancyhdr}

\usepackage[backend=biber,style=apa,uniquelist=false,uniquename=false,dashed=true,dateabbrev=false,eventdate=comp,hyperref=true,language=british,parentracker=true]{biblatex}

\DeclareNameFormat{labelname:poss}{% Based on labelname from biblatex.def
	\nameparts{#1}% Not needed if using Biblatex 3.4
	\ifcase\value{uniquename}%
	\usebibmacro{name:family}{\namepartfamily}{\namepartgiven}{\namepartprefix}{\namepartsuffix}%
	\or
	\ifuseprefix
	{\usebibmacro{name:first-last}{\namepartfamily}{\namepartgiveni}{\namepartprefix}{\namepartsuffixi}}
	{\usebibmacro{name:first-last}{\namepartfamily}{\namepartgiveni}{\namepartprefixi}{\namepartsuffixi}}%
	\or
	\usebibmacro{name:first-last}{\namepartfamily}{\namepartgiven}{\namepartprefix}{\namepartsuffix}%
	\fi
	\usebibmacro{name:andothers}%
	\ifnumequal{\value{listcount}}{\value{liststop}}{'s}{}}
\DeclareFieldFormat{shorthand:poss}{%
	\ifnameundef{labelname}{#1's}{#1}}
\DeclareFieldFormat{citetitle:poss}{\mkbibemph{#1}'s}
\DeclareFieldFormat{label:poss}{#1's}
\newrobustcmd*{\posscitealias}{%
	\AtNextCite{%
		\DeclareNameAlias{labelname}{labelname:poss}%
		\DeclareFieldAlias{shorthand}{shorthand:poss}%
		\DeclareFieldAlias{citetitle}{citetitle:poss}%
		\DeclareFieldAlias{label}{label:poss}}}
\newrobustcmd*{\posscite}{%
	\posscitealias%
	\textcite}
\newrobustcmd*{\Posscite}{\bibsentence\posscite}
\newrobustcmd*{\posscites}{%
	\posscitealias%
	\textcites}

\DeclareFieldFormat*{edition}{\nth{#1} ed.}

\newcommand{\noop}[1]{}

\usepackage{siunitx} % centering in tables

\def\yyy{%
	\bgroup\uccode`\~\expandafter`\string-%
	\uppercase{\egroup\edef~{\noexpand\text{\llap{\textendash}\relax}}}%
	\mathcode\expandafter`\string-"8000 }

\def\xxxl#1{%
	\bgroup\uccode`\~\expandafter`\string#1%
	\uppercase{\egroup\edef~{\noexpand\text{\noexpand\llap{\string#1}}}}%
	\mathcode\expandafter`\string#1"8000 }

\def\xxxr#1{%
	\bgroup\uccode`\~\expandafter`\string#1%
	\uppercase{\egroup\edef~{\noexpand\text{\noexpand\rlap{\string#1}}}}%
	\mathcode\expandafter`\string#1"8000 }

\usepackage{setspace}
\usepackage[bottom]{footmisc}
\usepackage[framemethod=tikz]{mdframed}

\usepackage{etoolbox}
\makeatletter
\patchcmd{\maketitle}{\@makefntext}{\fakecommand}{}{}
\patchcmd{\maketitle}{\rlap}{\hbox}{}{}
\patchcmd{\@maketitle}{\@author}{\hspace*{5pt}\@author}{}{}
\makeatother

\usepackage{comment}

\usepackage{changes}

\definechangesauthor[color=Green]{NA}

\definechangesauthor[color=Blue]{RWE}

\definechangesauthor[color=Red]{EFM}

\usepackage{todonotes}
\usepackage{abstract}
\usepackage{titling}

\usepackage[noblocks]{authblk}

\usepackage[pdftex,pdftitle={},pdfsubject={},%
pdfauthor={},citecolor=blue,colorlinks]{hyperref}
\hypersetup{
	colorlinks=true,%
	citecolor=blue,%
	filecolor=black,%
	linkcolor=black,%
	urlcolor=black}

\title{Does robotization affect job quality? Evidence from European regional labour markets\thanks{Corresponding author: Rudolf Winter-Ebmer, Department of Economics, Johannes Kepler University Linz, Alternberger Sta{\ss}e, 4040 Linz, e-mail: \href{mailto:rudolf.winterebmer@jku.at}{\texttt{rudolf.winterebmer@jku.at}}. Ant{\'o}n acknowledge funding from the Ram{\'o}n Areces Foundation (\nth{17} National Competition for Social Sciences Research Grants). Winter-Ebmer acknowledges funding from the Linz Institute for Technology (LIT) and the FWF. The authors thank comments from participants at the \nth{32} Annual Conference Society for the Advancement of Socio-Economics (July 2020, Amsterdam) and the \nth{32} Annual European Association for Evolutionary Political Economy (September 2020, Bilbao).}}

\pretitle{\vspace{-60pt}\begin{center}\Large\bfseries}
	\posttitle{\par\end{center}}

\predate{\begin{center}}
	\postdate{\end{center}\vskip -2em}
\makeatletter
\AtBeginDocument{%
	\newlength{\temp@x}%
	\newlength{\temp@y}%
	\newlength{\temp@w}%
	\newlength{\temp@h}%
	\def\my@coords#1#2#3#4{%
		\setlength{\temp@x}{#1}%
		\setlength{\temp@y}{#2}%
		\setlength{\temp@w}{#3}%
		\setlength{\temp@h}{#4}%
		\adjustlengths{}%
		\my@pdfliteral{\strip@pt\temp@x\space\strip@pt\temp@y\space\strip@pt\temp@w\space\strip@pt\temp@h\space re}}%
	\ifpdf
	\typeout{In PDF mode}%
	\def\my@pdfliteral#1{\pdfliteral page{#1}}% I don't know why % this command...
	\def\adjustlengths{}%
	\fi
	\ifxetex
	\def\my@pdfliteral #1{}% isn't equivalent to this one
	\def\adjustlengths{\setlength{\temp@h}{-\temp@h}\addtolength{\temp@y}{1in}\addtolength{\temp@x}{-1in}}%
	\fi%
	\def\Hy@colorlink#1{%
		\begingroup
		\ifHy@ocgcolorlinks
		\def\Hy@ocgcolor{#1}%
		\my@pdfliteral{q}%
		\my@pdfliteral{7 Tr}% Set text mode to clipping-only
		\else
		\HyColor@UseColor#1%
		\fi
	}%
	\def\Hy@endcolorlink{%
		\ifHy@ocgcolorlinks%
		\my@pdfliteral{/OC/OCPrint BDC}%
		\my@coords{0pt}{0pt}{\pdfpagewidth}{\pdfpageheight}%
		\my@pdfliteral{F}% Fill clipping path (the url's text) with
		\my@pdfliteral{EMC/OC/OCView BDC}%
		\begingroup%
		\expandafter\HyColor@UseColor\Hy@ocgcolor%
		\my@coords{0pt}{0pt}{\pdfpagewidth}{\pdfpageheight}%
		\my@pdfliteral{F}% Fill clipping path (the url's text)
		\endgroup%
		\my@pdfliteral{EMC}%
		\my@pdfliteral{0 Tr}% Reset text to normal mode
		\my@pdfliteral{Q}%
		\fi
		\endgroup
	}%
}
\makeatother

\DeclareLabeldate{%
	\field{date}
	\field{eventdate}
	\field{origdate}
	\field{urldate}
	\field{pubstate}
	\literal{nodate}
}

\renewbibmacro*{addendum+pubstate}{%
	\printfield{addendum}%
	\iffieldequalstr{labeldatesource}{pubstate}{}
	{\newunit\newblock\printfield{pubstate}}}

\author[$\dag$]{Jos\'e-Ignacio~Antón}
\affil[$\dag$]{University of Salamanca}
\author[$\ddag$]{Enrique~Fern{\'a}ndez-Mac{\'i}as}
\affil[$\ddag$]{Joint Research Centre, European Commission}
\author[$\S$]{Rudolf~Winter-Ebmer}
\affil[$\S$]{Johannes Kepler University Linz, IHS, IZA, CReAM and CEPR}

\begin{document}
	
\date{First version: \nth{22} December 2020\\This version: \today}

\maketitle

\singlespacing	

\begin{abstract}
	\noindent Whereas there are recent papers on the effect of robot adoption on employment and wages, there is no evidence on how robots affect non-monetary working conditions. We explore the impact of robot adoption on several domains of non-monetary working conditions in Europe over the period 1995–2005 combining information from the World Robotics Survey and the European Working Conditions Survey. In order to deal with the possible endogeneity of robot deployment, we employ an instrumental variables strategy, using the robot exposure by sector in other developed countries as an instrument. Our results indicate that robotization has a negative impact on the quality of work in the dimension of work intensity and no relevant impact on the domains of physical environment or skills and discretion.
	\vskip 0.5em

	\vskip 0.5em
	
\end{abstract}
	
\singlespacing	
\newpage
\section{Introduction}\label{Section 1}

Automation represents a major shaping force of today's labour markets, contributing to rising living standards \parencite{Atack2019, Autor2015, Autor2018}, but also being considered a relevant source of anxiety for citizens:  75\% of Europeans see technological progress as a phenomenon threatening their job prospects \parencite{Eucom2017a}. While there is increasing empirical evidence showing positive effects of robot adoption on productivity \parencite{Dauth2021,Graetz2018,Jungmittag2019,Kromann2020}, research on the impact of this technology on the labour market is mainly limited to employment and wages \parencite{Acemoglu2020a, Anton2022, Borjas2019, Chiacchio2018, Dahlin2019, Dauth2021, Graetz2018, Jager2016, Klenert2022, Koch2021, DeVries2020}.

The aim of this paper is to  explore whether the increase of density of industrial robots in Europe affects working conditions, in general. This is relevant for two reasons. In the first place, workers do care for working conditions. Workers are willing to trade money for improvements in other domains in the sense of compensating differentials  \parencite{Clark2015, Bustillo2011a, Maestas2018}, even if labour market imperfections and job rationing do not guarantee that the market compensates such attributes according to workers' preferences \parencite{Bonhomme2009}. Working conditions are one of  two most important concerns for European citizens \parencite{Eucom2017b}: more than a half of them reported that the quality of work has worsened during the last years.

Morever, other than business cycle fluctuations or changes in bargaining conditions, the introduction of (industrial) robots should in principle modify working conditions directly, but it is unclear in which direction. The main applications of this technology, whose adoption took off in the late 80s and early 90s, are handling operations and machine tending, assembling and disassembling and the sequential addition of standardized interchangeable parts to a complex product (e.g., a car) \parencite{IFR2017a}. There are several channels through which they could impact on different areas of working conditions, essentially, by modifying the tasks performed by employees.

First, in relation to work intensity, one should bear in mind that robots are centrally controlled machines with which workers have to interact. 	 They can therefore intensify work rhythms and labour effort if they make monitoring workers' performance easier---as suggested by \textcite{Brown2010} or \textcite{Weil2014} for certain technologies---and make employees' tasks more dependent on the pace of work of machines.

Second, regarding the physical environment of the workplace, robots replace certain tasks, which tend to be the most repetitive and heavy, thus contributing positively to job quality in terms of health and safety. The introduction of technology can prevent occupational accidents and diseases, but may also mean the appearance of new risks associated to malfunctions (which result in incidents like collisions and unexpected movements) \parencite{Vautrin1986}.

Third, robots require standardisation of processes, which can affect workers by reducing their autonomy, making employees' tasks more dependent on them.  Furthermore, robot adoption might also result in workers' relocation within the same firm. Employees whose tasks this technology assumes can climb the occupational ladder and take new tasks that require higher skill levels than the previous one \parencite{Dauth2021}; these tasks could, therefore, be more meaningful and fulfilling.
	
Furthermore, the raise in productivity due to robot adoption might also result in a wider space for improving not only wages but also other conditions at the workplace, even if they are costly for employers \parencite{Clark2015}; gains which, in turn, are shaped by the possibility of monitoring workers' performance \parencite{Bartling2012}. Therefore, the expected impact of robotization on job quality is quite ambiguous. It might alter the tasks performed by the workforce and, in principle, this effect can be more direct on those workers  complementary to robots than those who are potentially replaceable. Nevertheless, the presence of robots might also create pressure on the work carried out by other workers, who might have to take up or modify the way they perform their (new) tasks due to the introduction of the technology, particularly, if it enlarges the possibilities of employee surveillance.

We combine sector- and industry-level data on robots with several European-level surveys on  working conditions that allow us to analyse how the increase in robot density affects working conditions at the local labour market. We use composite indices of job quality previously employed in the literature and use Ordinary Least Squares (OLS) in changes over the period 1995-2005. As there might be reverse causality or a problem with missing variables, we resort to instrumental variables (IV) techniques based on  sector-level trends in robot adoption in other leading countries, such as South Korea, Sweden, Switzerland, or Australia. Whatever method we use, we find that the increase in robot adoption across Europe had a negative effect on job quality in work intensity but does not have any effect on other aspects of job quality, like physical job environment and skills and discretion of workers on the job.

Robots applied in the industrial production process are not so widespread as computers and their applications have little to do with the ones of artificial intelligence. Therefore, it is debatable to which extent the deepening in the adoption of this technology represents a qualitative change \parentext{as one could more easily argue, for instance, regarding artificial intelligence \parencite{Acemoglu2020b, Fernandez2021}}.\footnote{These robots carry out physical tasks involving the moving and precise manipulation of objects within standardized industrial processes. These robots are not anthropomorphic, many of them resemble arms. They typically end in an effector (which might look like a human hand) that often carries out a precise manipulation task. Usually, they remain within a predefined and limited space and are constrained to a very particular task. The applications of these robots, mentioned above, suggest that they do not represent a radical departure from the long-term process of industrial automation but its latest iteration. Most of these industrial robots perform essentially the same type of operations as previous mechanization and automation technologies, replacing labor input in routine tasks that involve physical strength and dexterity.} As a consequence, the impact of robots might not be the same as in the case of other technologies previous literature has analysed thoroughly \parentext{see, \textcite{Acemoglu2011}, \textcite{Autor2015}, \textcite{Autor2018}, \textcite{Barbieri2019}, \textcite{Fernandez2016} and \textcite{Jerbashian2019}}.

With our research we contribute to several recent discussions. There is a growing literature on the impact of robotization on employment and wages, which is far from having reached a consensus. While several studies  for the US suggest a negative impact on employment and wages \parencite{Acemoglu2020b,Borjas2019,Dahlin2019}, the effects are not so clear for other economies. Those negative effects only seem to apply to manufacturing in Germany \parencite{Dauth2021}, while the work of \textcite{Chiacchio2018} for six EU countries only shows a detrimental impact on employment, but not for wages. The pioneering study of \textcite{Graetz2018}, including a wide set of developed countries, identifies a positive effect on wages and a neutral effect on employment, whereas \textcite{Klenert2022}, which extend the period of analysis and limit their exploration to manufacturing, even point out  a positive contribution of robots to employment growth.\footnote{See also \textcite{Bekhtiar2021} and \textcite{Fernandez2021} on problems with industry-level data.} Similarly, the cross-country studies of \textcite{Carbonero2018} and \textcite{Debacker2018} draw mixed conclusions on the impact of robots on job creation. Firm-level studies, on the other hand, find a positive association between the introduction of robots and employment growth in France \parencite{Domini2021} and Spain \parencite{Koch2021}.

So far, there is only limited research on other non-monetary aspects of job quality. Closer to our topic are recent papers by \textcite{Gihleb2020, Gunadi2021, Lerch2020}. While \textcite{Lerch2020} is mostly concerned with labor market effects of robots, he also presents some evidence that robot adoption might be correlated with the prevalence of health problems or admissions to hospitals for displaced workers. \textcite{Gihleb2020} and \textcite{Gunadi2021} use data on industrial robot penetration and find that increased robot use is reducing work-related injuries and improving health in the US, finding a negative relationship. The work of \textcite{Gihleb2020}, which is contemporaneous to ours also explores the effect on work intensity in Germany, without identifying any impact. Nevertheless, it is worth mentioning that their data and set-up only allows them to explore between-occupation change in the latter variable \parentext{while a large part of the variation in what workers do takes place within occupations \parencite{Freeman2020,Maier2021}}. Our study is therefore the first to explore the impact of robots on different areas of working conditions and working conditions in general. Moreover, we can differentiate between different aspects of working conditions and with a focus on the European labour market.

We also contribute to the discussion about changes in working conditions. \textcite{Fernandez2015} show that job quality in the EU was remarkably stable before and after the financial crises with some increase in job quality in the European periphery. \textcite{Green2013} look at different components of working conditions and find the component of work intensity and---to some extent--working time quality to improve in Europe. Moreover, they study the dispersion of these measures across groups and across time. \textcite{Bryson2013} investigate the impact of organisational changes and trade unions on working conditions, whereas \textcite{Cottini2013} explore the consequences of working conditions on mental health. Closer to our topic are studies relating changes in computer use with working conditions: \textcite{Menon2020} report that computerization has no large effects on working conditions in general, there is even a mild positive effect on job discretion. \textcite{Green2001} in an earlier study show that computer use leads to an intensification of the workpace.\footnote{There is a large literature on job satisfaction or happiness as general indicators of working conditions \parentext{see \textcite{Clark1994} or \textcite{Clark2005} for early references}. These indicators may lack comparability as they may also comprise differences in expectations \parencite{Osterman2013}; they are general indicators and there is no research linking these indicators to robots.} In our study we extend this analysis with a closer look at the impact of robotization on working conditions in general.

Following this introduction, the rest of the paper unfolds as follows. Section 2 describes the databases employed in the analysis and outlines the identification strategy used in the econometric analysis. We present the main results of the paper in section 3 and the \nth{4} and last section summarizes and discusses the main conclusions of the paper.

\section{Data and methods}\label{Section 2}

\subsection{Data}\label{Subsection 2.1}

\textit{Robots}. In order to assess the effect of robotization on the working conditions of European workers, we combine several databases that contain information on robot adoption and job attributes. Our first source is the World Robotics 2017 edition, a dataset administered by the International Federation of Robotics \parencite{IFR2017b}, the main association of manufacturers of robots worldwide. It comprises information on industrial robot stocks and deliveries by country and sector of activity all over the world from 1993 to 2016. As mentioned above, the robots included in the \textcite{IFR2017b} consist of industrial machinery, digitally controlled, mainly aimed at handling operations and machine tending, welding and soldering and assembling and disassembling. In terms of accounting, these robots are part of non-information and communication technology capital, with the exception of their associated software needed to manage them.\footnote{\textcite{IFR2017b} provides robot figures by industry according to the International Standard Industrial Classification of All Economic Activities, Revision 4, which is largely compatible with the Statistical Classification of Economic Activities in the European Community, Revision 2 (NACE Rev. 2).}

The IFR basically constructs a series of robot stocks on the basis of deliveries, using a perpetual-inventory approach and a 12-year depreciation. This is a more reliable approach---as compared to using stocks---, since the association of robot producers  controls those inflows directly. As the distribution of robots is missing in some years and countries, we impute initial unspecified stocks or deliveries on the basis of the distribution by industry in the three closest years to the period of interest with specified information.\footnote{This process is very similar to the one followed by \textcite{Graetz2018}. They use the total number of specified deliveries for imputation (instead of the three closest years). Our series are virtually identical. For more details on the imputation procedure, see the supplementary material of \textcite{Fernandez2021}.}

\textit{Working Conditions}. We use the \nth{2}, \nth{3} and \nth{4} waves the European Working Conditions Survey (EWCS), carried out in 5-year intervals by the European Foundation for the Improvement of Living Conditions (Eurofound) \parencite{Eurofound2018b}, 1995, 2000 and 2005.\footnote{We do not include the first wave of the EWCS, because of limited coverage of countries, the substantial smaller number of variables due to job quality, and the absence of robot data prior 1993 in the IFR database.} There are two additional waves (2010 and 2015), but we focus on the period 1995--2005 in order to avoid  confounding effects of the Great Recession, which had a markedly different impact across countries and regions.\footnote{The Great Recession implied large asymmetric shocks across European countries and even regions within the same country. For instance, the length of the crisis in Germany or Sweden is roughly a year, while Spain, Portugal or Greece were still much below the pre-crisis GDP per capita and employment levels in 2015. Previous literature highlights the relevance of this contraction for working conditions, so its consideration could be problematic, particularly, in terms of work intensity in manufacturing. E.g., in Spain, many indicators improved because of the lack of demand \parencite{Bustillo2011c} and there were declines in some dimensions in other countries \parencite{Fernandez2015, Eurofound2021, Vaughan2011}.} For our sample countries, the number of robots per thousand workers rose from  0.800 in 1995 to almost 2.200 in 2005; with only a slight increase to  2.750 in 2015. Finally, as there is less variation in the latter period, our proposed IV strategy does not produce a strong first stage. In any case, we present the results for this second period in the annex and comment on them in the next section.

The EWCS represents the most comprehensive database for the analysis of non-monetary working conditions across Europe on a comparative perspective, covering the European Union (EU) members, several accession countries and other states like Norway and Switzerland. We focus on the 12 EU countries  with the highest ratio of robots per worker during the analysed period.\footnote{The list of countries includes Austria, Belgium, Denmark, Finland, France, Germany, Italy, the Netherlands, Portugal, Spain, Sweden and the United Kindgom.}

Each wave includes a minimum of 1,000 interviewees in each country and year. As robot technology is mainly used in manufacturing, we  focus on  privately salaried workers employed in mining and quarrying and the secondary sector (manufacturing, electricity, gas  water supply and construction), which concentrates more than 90\% of these types of robots. Unfortunately, there is no further disaggregation of these industries. This leaves us a sample of 7,764 workers that we collapse in order to obtain the region-level outcomes and covariates. The EWCS contains a rich set of variables covering different dimensions of working conditions; we describe that in the next subsection.

\textit{Control Variables}. We use the European Union Labour Force Survey (EU-LFS) \parencite{Eurostat2019} and due to missing regional information for Germany the European Community Household Panel (ECHP) for the years 1995 and 2000 \parencite{Eurostat2003}.\footnote{Detailed information by region and industry sometimes was only available through an ad hoc request to the Eurostat User Support \parencite{Eurostat2020a}. Moreover, sectors have to be reclassified from NACE Rev. 1.1 to NACE Rev. 2.} Changes in information, communication and technology (ICT) capital stock per worker are  from the EU KLEMS \parencite{Stehrer2019}, data for Chinese imports come from  the United Nations International Trade Statistics Database, to which we access through the World Integrated Trade Solution (WITS) \parencite{Worldbank2020} following \textcite{Autor2013b}.\footnote{Previous literature finds relevant negative effects of exposure to Chinese imports not only on employment and wages \parencite{Autor2016}, but also on health outcomes \parencite{Lang2019}.}

We construct instrumental variables from the Korean Industrial Productivity Database (KIP), provided by the \textcite{KPC2015}, labour force statistics from Australia and information from Eurostat for Switzerland.

Our database on working conditions, the EWCS, does not contain detailed information on sectors of activity (only some large industry groups are available), but it is representative by region. In the fashion of previous literature \parencite{Acemoglu2020a, Dauth2021}, we perform the analysis at such a level. In order to calculate the change in robot exposure by region, assuming that the distribution of the change in robot stocks by region over a certain period of interest depends on the distribution of employment at the beginning of the period, we combine detailed sector-level data by country on robots and region-level employment data by industry obtained from several ad hoc requests to the Eurostat User Support \parencite{Eurostat2020a}.\footnote{We consider 20 sectors of activity (in the nomenclature of NACE Rev. 2) we are able to match in our robot and employment data: agriculture, hunting and forestry; fishing (A); mining and quarrying (B); food products and beverages; tobacco products (C10--12); textiles, leather, wearing apparel (C13--15); wood and wood products (including furniture) (C16); paper and paper products; publishing and printing (C17--18); chemical products, pharmaceuticals, cosmetics, unspecified chemical, petroleum products (C19--21); rubber and plastical products (C22); glass, ceramics, stone, mineral products not elsewhere classified (C23); basic metals (C24); metal products (except machinery and equipment) (C25); electrical/electronics (C26--27), industrial machinery (C28); automotive (C29); other transport equipment (C30), other manufacturing branches (C32); electricity, gas and water supply (D, E), construction (F), education, research and development (P) and others.	The questionnaires of the EU-LFS effectively collects this detailed information on the distribution of the labour force by region and industry, but the anonymized microdata does not disclose it because of confidentiality reasons and we access it to several tailored petitions.} We provide further details on the construction of the variation in the robot exposure by region in the next subsection.

\subsection{Methodology}\label{Subsection 2.2}

As mentioned above, our identification strategy exploits the regional variation in the increase in the adoption of robots. Following the strategy proposed by \textcite{Acemoglu2020a}, we compute the change in the exposure to robotization by region assuming that the robot inflows during a certain interval of time follows the distribution of employment in the initial period. Our geographical units of analysis mainly correspond to the Nomenclature of Territorial Units for Statistics at the second level (NUTS 2), although in some cases, because of the existence of administrative changes in the boundaries of NUTS we cannot trace over time, we make use of larger geographical units. As a result, we are able to trace 80 regions over the period 1995--2005. Given the very low mobility across NUTS 2 in Europe \parencite{Gakova2008,Janiak2008}, we can consider our regions as reasonably closed labour markets, in the sense that it is not likely that robot adoption results in relevant outflows from regions with high deployment to others with low adoption of the technology.\footnote{The units considered in the analysis correspond to NUTS 2, NUTS 1, or even larger territories (e.g., combinations of several ones because of the changes in the nomenclature over time). The eventual different size of the regions is not problematic as long as we consider enough large units where out-migration is no issue.} Another problem that might arise is a sort of sample selection bias. We concentrate our analysis on workers in these regions in the years 1995, 2000 and 2005. If the exposure to robots would change or reduce the workforce considerably, we would be in trouble, comparing working conditions of those in the region before the advent of robots with the working conditions of those still employed in the region after the exposure to robot adoption. While \textcite{Acemoglu2020a} do find a negative impact of robots on employment in the US, studies for Europe do not find such effects \parencite{Dauth2021,Anton2022}. In addition, we use additional variables to control for changes in the composition of the workforce.

The main variable of interest in our analysis is the increase in robot exposure (RE), which we define as the change in the number of robots during a certain period $t$ in a region $r$ divided by the number of workers in the region at the beginning of the period, that is,
\begin{equation}
\label{eq1}
\Delta{\mathit{RE}}_{rt} = \frac{1}{L_{rt}}\sum\limits_j{{\frac{L_{rjt}}{L_{jt}}}{\Delta R_{jt}}}
\end{equation}
where $R_{jt}$ represents the change in robot stocks in sector $j$ in the country where the region is located over period $t$; $L_{rt}$, the initial number of workers in the region at the beginning of the period of interest (i.e., 1995 for the first difference, 1995--2005, and 2000 for the second time lapse, 2000--2005); $L_{jt}$ denotes the employment figures in industry $j$ in region $r$ in the initial year and $L_{rjt}$, the number of workers in region $r$ in industry $j$ at the same moment of time. In this fashion, we attribute to each region a change in the stock of robots according to the share of employment in this sector in the initial period.{\footnote{Regrettably, the EU-LFS is not available for most of the countries before 1995, which prevents us from using lagged employment shares. Under this circumstance, one  cannot compute the so-called Rotemberg weights in this shift-share design \parencite{Goldsmith2020}. Within this framework, the lagged share of employment would work as an instrument for the change in robot density and the Rotemberg weight would capture how much bias the overall estimate would have if a given instrument (industry) were biased by certain percentage. Nevertheless, in our set-up, the procedure should be less problematic than in the canonical applications of \textcite{Goldsmith2020}, because we know with certainty the robot exposure for all our countries separately and, then, for each country, we make use of the shift-share approach in order to compute the regional exposure. Furthermore, following \textcite{Dottori2021}, who find very large Rotemberg weights for some industries,  we will show below that our results are robust to the use of instruments based on the robotization in countries with different relevance of the automotive sector.}

In order to explore the impact of robot adoption on working conditions, we estimate the following equation:
\begin{equation}
\label{eq2}
\Delta Y_{rt}= \beta_{0} + \Delta RE_{rt}\beta_{1} + Z{^{\prime}_{rt}}\beta_{2} + \varepsilon_{rt}
\end{equation}
$\Delta Y_{rt}$  denotes the change in the average job quality indicator of region $r$ over the period $t$. $Z{^{\prime}_{rt}}$ is a set of start-of-the-period regional control variables, very similar to those considered by \textcite{Acemoglu2020a} and \textcite{Autor2013b}, including the share of employment in mining and quarrying and the secondary sector in the region (to which we refer as the share of industry for brevity), population (in logs), share of females, age structure of the workforce, the share of population with middle or high education, the average routine task intensity (RTI) \parencite{Autor2013a,Goos2014,Mahutga2018,Schmidpeter2021} and the average offshoreability risk \parencite{Blinder2013,Mahutga2018}.\footnote{The aim of controlling for the initial values of RTI and offshoreability is to rule out that other sources of labour market changes due to technological changes different from robot adoption  might conflate with the latter. The RTI intends to capture to which extent an occupation is routine-task intensive. The logic behind the RTI measure is that automation is more likely to affect routine, manual, non-interactive job tasks. Likewise, the offshoreability risk index tries to measure the degree to which a certain occupation might be outsourced to a remote location.} Note that using an econometric specification with both outcome and the treatment variable in changes, we control for regional time-constant heterogeneity. Given that we pool two five-year differences, we include time fixed effects covering each of those periods. Second, we add a geographical dummy for core-periphery countries to capture group-of-countries-specific time trends.\footnote{We are unable to include country dummies given that some states only contain one traceable region because of their size or changes in the boundaries of NUTS2.}. Finally, it is possible that some changes in working conditions  have to do with changes in the labour force composition. In order to mitigate this selection effect, we control for the changes in the share of female workers, the proportion of workers with medium education, the proportion of workers with high education and the share of workers aged less than 30 years old and aged 50 years old or more employed in the region in the industries considered in the analysis.

Similar to previous analyses of the impact of robotization on employment or wages \parencite{Acemoglu2020a, Dauth2021}), there is the possibility of reverse causation. In these studies it may well be that robot adoption is caused by developments on the labour market, like the availability of suitable workers or a fast rising wage in the respective sector. In our case, reverse causation could occur for similar considerations: Since working conditions can also be indirect cost components (slower work pace or costs for accident avoidance) or have an impact on labour supply with respect to a specific industry, reverse causation could apply. Given our use of non-monetary working conditions, the argument for reverse causation is less strong as in the case of wages. Still, we use the same strategy as \textcite{Acemoglu2020a} and \textcite{Dauth2021}, who instrument the adoption of robots by the trends in other developed countries.\footnote{It is worth mentioning that previous studies using robot data from the IFR to explore the impact of this technology on labour market outcomes find very close OLS and IV estimates \parencite{Acemoglu2020a, Dauth2021, Graetz2018}, thus suggesting little evidence of endogeneity in the first place.} Given  our  focus on European Union countries, we look at the patterns of robotization by sector in South Korea, one of the world leaders in the adoption of this technology \parencite{IFR2017a, UNCTAD2017}, in order to build our IV. Considering the size of this economy and its limited integration with EU countries (compared to other member states), it is not likely that Korean industry-level developments trigger any relevant general-equilibrium effects. The exclusion restriction of the IV strategy requires that the instrument (robotization in Korean industries) has no impact on European working conditions over and above its indirect impact via robotization in Europe. We strongly believe that this is, indeed, the case.  We also build on data from Sweden (the pioneer in robot adoption in Europe) and Australia and Swizerland (two developed economies outside the European Union) in order to check the robustness of our results using alternative instruments.

We can express the increase in robot exposure as a function of the importance of each industry in the region and the average increase in robot density per worker at the national level. In order to build our IV, we consider the increase in robot exposure per worker in each of our third countries instead of the variable corresponding to each European Union country, obtaining the following expression:
\begin{equation}
\label{eq3}
\Delta{\mathit{RE}}^{k}_{rt} = \frac{1}{L_{rt}}\sum\limits_j{L_{rjt}\frac{\Delta R_{jt}^{k}}{L_{jt}^{k}}}
\end{equation}
where the superindex $k$ denotes the third country used for building the IV (South Korea, Sweden, or Australia and Switzerland). Our IV is relevant, with an \textit{F}-statistic between 40 and 80 in different  econometric specifications, using clustered standard errors at the regional level.\footnote{In the case of our third IV specification, in order to get an strong enough first stage, we include the change in robot exposure per worker based on Australian data and the same variable based on Swiss data at the same time.}

In the construction of our variable \enquote{change in robot exposure}, we proceed as follows. The increase in robot density in a certain region draws on the change in robot stock and the employment at the beginning of the period of interest at the regional level. Therefore, it requires the number of robots and workers in each sector at the two-digit level. The former variable comes from the IFR data, whereas the latter comes from the population estimates of the EU-LFS (which can be obtained through the sampling weights). We obtain the number of workers (or any other variable) in each industry at the required level from the EU-LFS through the Eurostat user support. As a result, we are able to construct a variable capturing the change in exposure to robot adoption at the regional level.

In order to build  changes in ICT capital stock per worker and in the exposure to Chinese imports, we depart from sector-level data and follow a similar procedure to the one applied to robots based on the initial distribution of employment, considering roughly the same industry classification as in the case of robots and even a more detailed one regarding Chinese imports.

Our measures of working conditions, based on the EWCS and developed by Eurofound and their collaborators \parentext{see, e.g., \textcite{Eurofound2012,Eurofound2015,Eurofound2019}, \textcite{Fernandez2015}, \textcite{Green2013}, \textcite{Menon2020}, and \textcite{Bustillo2011a}}, comprise three dimensions: work intensity, physical environment and skills and discretion. We reformulate these indicators in order to ensure that the variables of interest are available in the three waves of the survey. It is relevant to highlight that this set of indicators privileges the inclusion of \enquote{objective} rather than \enquote{subjective} variables when possible in order to minimize the effect of adaptation, adjustment, or cognitive dissonance \parencite{Bhave2013, Bowling2005, Bustillo2011b,Pugh2011,Ritter2016}.\footnote{There are other dimensions of job quality proposed by the Eurofound: working-time quality, social environment, and prospects. We do not consider them in the analysis because most of the key variables that integrate them are not available in all waves. The interested reader can find details on all the methodological issues related to the operationalisation of the job quality indicators in the literature \parencite{Eurofound2012, Eurofound2015, Eurofound2019, Fernandez2015, Green2013, Menon2020, Bustillo2011a}. The procedures to construct them unfold as follows. We employ the raw variables (i.e., the questions available in the survey) to define all the dimensions of job quality, whereby a higher value of the indicator means a better job, transforming all the items using a min-max normalisation between 0 and 100. For instance, if a variable runs from 1 (best value) to 5 (worst value), we compute $100\cdot(5-\text{value of the variable})/(5-1)$. Each subdimension provides an average score of its component variables and, in turn, each dimension is the result of the arithmetic mean of the scores of the sub-dimensions. In the same fashion as the bulk of the recent research using these kinds of measures \parentext{see, e.g., \textcite{Eurostat2019}}, each variable receives the same weight within each sub-dimension and we assign the same importance to each sub-dimension when computing the score for each dimension. Sensitivity analyses in \textcite{Bustillo2011a} suggest that these composite measures are robust to the use of different weighting schemes because there is a high positive correlation between the outcomes in different dimensions.}

The index of work intensity comprises two sub-dimensions, quantitative demands and pace determinants and interdependency. The first sub-dimension builds on three variables, pace of work (high speed), tight deadlines and time pressure, while our indicator of pace determinants and interdependency considers how interviewee's work depends on colleagues, customer demands, production targets, machine speed and bosses.

Job quality in physical environment considers three domains: ambient risks (vibrations, noise, high temperatures and low temperatures), biological and chemical risks (exposition to fumes and vapours and chemicals) and posture-related risks (tiring positions, heavy loads and repetitive movements).

Finally, the quality of work in terms of skills and discretion comprises three sub-dimensions: cognitive tasks (carrying out complex tasks and working with computers, smartphones, laptops, etc.), decision latitude (control the order of tasks, speed of work, methods of work and timing of breaks) and training (receiving training provided by the employer and the possibility of learning new things).

Following the previous literature \parentext{see, e.g., \textcite{Eurofound2019}}, we combine these variables, most of them of an ordinal nature, in order to define indicators of job quality in each of the dimensions and sub-dimensions in a positive sense---i.e., the higher the measure, the higher the well-being---and using a 0--100 scale. For instance, the attribute \textit{vibrations} receives the highest score when the workers are never exposed to this sort of workplace risk. Each variable receives the same weight within each sub-dimension and  we compute the arithmetic average of these sub-domains in order to again obtain  a score between 0 and 100 for our index of job quality in work intensity.\footnote{The sensitivity analyses presented by \textcite{Bustillo2011a} suggest that the composite measures of these dimensions are quite robust to the use of different weighting schemes because there is a high positive correlation between the outcomes in different domains.} Although, as argued above, our indicators of job quality draw on objective working conditions when possible (e.g., temperature level instead of satisfaction with the workplace temperature), in order to assess the robustness of our findings we also look at the effects of robot adoption on three subjective binary variables: the workers' self-awareness of the impact of work on their health and self-perceived work-related stress and anxiety.\footnote{The exact wording of the questions is \enquote{Does your work affect your health, or not?} and \enquote{How does it affect your health?}, respectively. In the latter question, the interviewer offers several possible answers, among them, stress and anxiety. It allows multiple answers.}

Our left-hand-side variable is a regional average of the indicator of interest using sampling weights (that the EWCS calculates  from the EU-LFS estimates) with the aim of making it representative of the corresponding population.

\section{Results}\label{Section 3}

Table~\ref{Table 1} displays  descriptive statistics of the dependent variables and covariates of our database, containing 80 European regions. We present the figures for the three mentioned dimensions (work intensity, physical environment and skills and discretion) and the two sub-domains composing work intensity. The evolution of these variables over time does not seem to follow a clear pattern. The number of robots per worker by region multiplies by more than 2.5 from 1995 to 2005. Figure~\ref{Figure 1} plots the correlation between 5-year changes in robot exposure and changes in job quality by dimension over the period 1995--2005. The graphs suggest a negative correlation in the case of work intensity, a somewhat weaker negative one with physical environment and a slightly positive one with respect to skills and discretion.

We present the main results of our analysis of the effects of robot adoption on work intensity, physical environment and skills and discretion in Tables~\ref{Table 2} and \ref{Table 3}, respectively.\footnote{Note that the number of observation is 180, instead of 240, because we estimate models in changes.} In these tables, we display both  OLS and IV estimates, without and with controls for the change in the share of workers of different characteristics in the working population in the region.
The relevant \textit{F}-statistic of the first stage is well above 50, pointing out to the relevance of our IV. We present the complete details on the first stage in Table~\ref{Table A1} in the Appendix.

Table~\ref{Table 2} shows that the adoption of robots reduces job quality with respect to  work intensity. All four estimates are very consistent, columns (1) and (2) excluding or including variables for compositional change in the workforce show an effect of $-4.5$, whereas the IV results are somewhat higher at $-5.2$--$-5.6$; the statistical indistinguishability between OLS and IV results indicates no big relevance for endogeneity. The quantitative result means that an increase in robot adoption of one unit (which is around one standard deviation in 2000) increases work intensity by 4--5 units (60--80 percent of a standard deviation in 2000). In other words, the increase in robots between 1995 and 2005 from $0.8--2.1$ per thousand workers led to an increase in work intensity of 5.6--7.3 points (87--114 percent of a standard deviation in 2000). These effects are rather large, but comparable to those of \textcite{Menon2020} in size: They calculate the effect of computers on working conditions in the European Union, finding negative but insignificant coefficients for the impact of computer use on work intensity, but a positive impact of computer use on work quality in terms of work discretion.

Table~\ref{Table 3} reports similar estimations for working conditions in terms of physical environment and skills and discretion. Panel A of the table refers to physical environment and Panel B to skills and discretion. Here, the effects are smaller, mostly negative (i.e., reducing job quality) and insignificant. This refers to both OLS and IV results: physical environment and skills and discretion are not impacted by the adoption of robots.

We have seen that there is a negative effect of robotization on job quality, but only in the dimension of work intensity, not in physical environment and skills and discretion. In Table~\ref{Table 4} we further proceed by looking at the sub-domains of work intensity, quantitative demands and pace and interdependence. Again, we present OLS and IV coefficients, which are fairly consistent. Both dimensions of work intensity are negatively related with robotization. The impact on the sub-dimension of quantitative demands is considerably  stronger than in the whole job quality dimension, while the effect in the case of pace and interdependency is somewhat weaker, but still statistically significant.

So far, we can say that work intensity increased for workers that were employed in   mining and quarrying and the manufacturing sector that adopted robots most intensively during this period. The results for industry might not have implications on working conditions in the whole economy if robot adoption there results in a displacement of workers from these economic activities to the rest of the economy. Drawing on EU-LFS data (which includes the reference population for employment rates and provides the framework for the sampling weights of the EWCS) and using the same identification strategy and taxonomy of regions as with working conditions, we test whether robot adoption affects the share of employed working-age people in industry, the rest of sectors and the whole economy. The results of this exercise, shown in Table~\ref{Table 5}, rule out this possibility.

The effect of robotization on wages should  also be interesting for interpreting the relevance of our results. Higher monetary compensation might compensate for worse working conditions. Unfortunately, our database does not allow testing for that mechanism in a similar fashion as in the domains considered here (this information is only available in a proper way since the fifth wave of the EWCS). Nevertheless, the work of \textcite{Chiacchio2018}, which considers a quite similar sample of countries, employs the approach based on robot adoption and follows an specification analogous to ours, finds no effect on wages in industry and a non-robust impact on remuneration of total workers (the estimated coefficient is significant and negative in some cases and not statistically different from zero in others) for the period 1995--2007. Therefore, the absence of evidence of wage increases due to robot adoption in Europe over the period of interest suggests that the potential compensation mentioned above does not apply here.

In order to check the robustness of our main results, we perform several additional estimations whose results are presented in Table~\ref{Table 6}. In the first two columns, we test whether our results hold under the use of other instruments: in column (I) we use two countries outside the European Union not included in our sample of regions, Australia and Switzerland. Under this specification, the effect of robot adoption remains negative and significant. In column (II), we use the increase in robot penetration by sector in Sweden (one of the leaders in the adoption of this technology in Europe) in order to build our IV. In this case, we have to  exclude Sweden from the countries considered in the analysis. Our results are pretty similar to the ones reported under our original instrument based on South Korea.

In the third column, we include two additional controls that, though being potentially endogenous, have been shown to influence labour market outcomes: the increase in exposure to Chinese imports and the increase in the ICT capital per worker. Moreover, they could correlate with the adoption of robots. Results have to be taken with care, therefore. The estimates in column (III) show that the baseline results do not qualitatively change when adding these additional covariates, corroborating our main results.

The fourth one displays the results when our regressions are not weighted by initial regional employment. Although we believe that weighting is desirable given the difference sizes of the geographical units (some of which we have to merge), the main message of our analysis holds.

In order to test whether the presence of certain influential regions might be the main driver of our results, in the same fashion as \textcite{Gunadi2021}, we compute the parameter estimates of the effect of robotization on job quality leaving out one region at a time (Figure~\ref{Figure 2}). These results confirm the outcome of our main analysis.

The final robustness check  is a rough \enquote{falsification} test, where we look at the effects of the change in robot exposure per worker in the region on workers in agriculture, forestry and fishing and the services sector. Given that most of the robots are concentrated in manufacturing, we should expect a null or, at least, a much lower impact of robots on the job quality of workers employed there. If our results were based on other concurrent events---correlated with robot adoption---such a placebo might catch these concurrent events. As expected, and in contrast to such a hypothesis, there is no effect of robotization in this falsification exercise (column [V]).

We present similar robustness checks for the impact of robotization on job quality in the dimensions, physical environment and skills and discretion in the Annex (Tables~\ref{Table A2} and \ref{Table A3}). These results are very similar to those  presented in Table~\ref{Table 3} and do not show any effect of robotization on either the physical environment of the job or skills and discretion in the job.

Given the impact in the work intensity identified above, it is worth exploring whether the effects on this domain translate into negative health consequences (Table~\ref{Table 7}). Our results suggest that robot adoption raises the share of employees reporting that their work affects their health, with a one-unit shift in robot density elevating this magnitude by 6.6 percentage points. Specifically, this effect seems to confirm the positive impact  of robotization on the proportion of workers declaring work-related stress and anxiety problems: an increase of one robot per thousand workers raises the probability of reporting those conditions by 9.2 and 3.2 percentage points, respectively. These outcomes are not at odds with the ones reported by the works of \textcite{Gihleb2020}, which associates a more intensive use of this technology with higher rates of  mentally unhealthy days among the workforce, and \textcite{Lerch2020}, which suggest a rise in hospital admissions and take-up rates of disability benefits because of robot adoption.

Last, we refer to the results for the period 2005--2015 (Table~\ref{Table A2}), which, as mentioned in Subsection~\ref{Subsection 2.1}, is subject to certain threats to identification and, therefore, its interpretation requires caution. In this time interval, robot adoption does not exert any effect on any of the job quality dimensions. Apart from the limitations commented above, we outline several explanations. First, it is possible that the higher intensity of robot adoption in the first period implies that the main effects of this technology in terms of workplace reorganization might have been observed during the first one. For instance, it is possible that the main changes in terms of task reorganization might have already taken place.\footnote{In their analysis on Germany, \textcite{Dauth2021} find that robotization encourages firms to relocate incumbent employees to other tasks and reduce the hiring of young workers. It is reasonable to think that, after this more disruptive impact in the beginning, companies do not have to introduce strong changes in their functioning to the same extent when robot adoption reaches a certain threshold.} 

Secondly, this result, which is coherent with the more negative impact of robots on employment identified by some works before the Great Recession \parencite{Anton2022,Bekhtiar2021,Chiacchio2018}, aligns with the larger impact of robot adoption on productivity in Europe after 2007 compared to the pre-crisis period} \parencite{Jungmittag2019}. Given the link between productivity and remuneration packages (which includes non-pecuniary working conditions), the absence of a negative impact of robots on job quality in the second period would be consistent with those larger productivity increases.

\section{Conclusions}\label{section 4}

The impact of technology on the workplace, workers and their work environment  attracts a lot of concern among citizens and researchers in Social Sciences, alike. The adoption of industrial robots, even if not new, is one of the more visible realizations of such technological changes. While there are a relevant number of studies concerned with the impact of this technology on employment and wages, ours is the first comprehensive study on the impact of robotization on working conditions in Europe.

We employ data from the European Working Conditions Survey and instrumental variables techniques in order to explore how a more intense adoption of this technology shapes job quality in regional labour markets. Over the period 1995 to 2005 an increase in robots used in industry led to worse working conditions with respect to tougher work intensity, but there are no effects on other working conditions, like physical environment of the job or skills and discretion in the job. Negative results for anxiety and stress on the job confirm our analysis. While robots are substituting for arduous---repetitive, heavy or fatiguing---tasks, their precision and predictability and standardization may lead to an increase in work intensity.

While work intensity, indeed, increased for workers in the manufacturing sector, which was instrumental in  the adoption of these robots, structural change---out of these sectors---could make a final assessment on total working conditions impossible. Additional evidence shows that, first, there is no displacement of workers out of manufacturing and, second, there are no changes in working conditions in the service sector. While we do find a consistently negative effect of robotization on working conditions in the period 1995--2005, where robot introduction exploded, there is no such effect in the period after the Great Recession: the reasons behind may be a previous adaptation to a new situation as well as generally more volatile employment and working conditions across European regions after the economic crisis.

\clearpage
\singlespacing
\printbibliography

\clearpage

\section*{Figures and tables}

% Figure 1: Job quality of work intensity and robot exposure by dimension (5-year differences, 1995--2005).

\begin{singlespace}
	\begin{figure}[htbp]
		\caption{Job quality index of work intensity and robot exposure (5-year differences, 1995--2005)}
		\centering \includegraphics[width=0.9\textwidth]{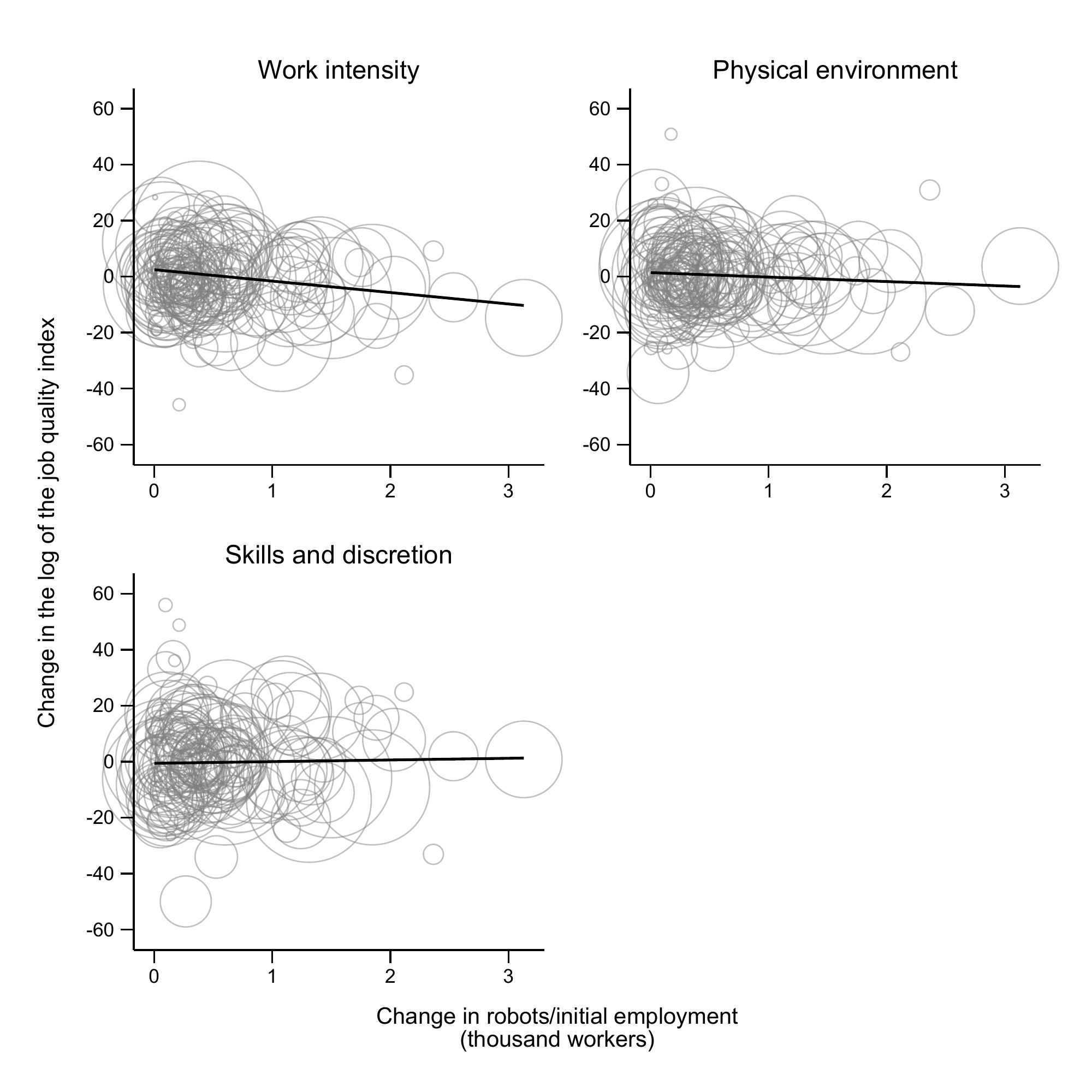} \\
		\justifying
		\footnotesize
		\noindent\textit{Note:} Observations weighted by the number of workers in the region at the beginning of the period.\\
		\noindent\textit{Source}: Authors' analysis from EWCS, EU-LFS and IFR.	
		\label{Figure 1}
	\end{figure}
\end{singlespace}

\clearpage

% Figure 2: Leave-one-out results

\begin{singlespace}
	\begin{figure}[htbp]
		\caption{Robustness checks: leave-one out test (5-year differences, 1995--2005)}
		\centering \includegraphics[width=0.9\textwidth]{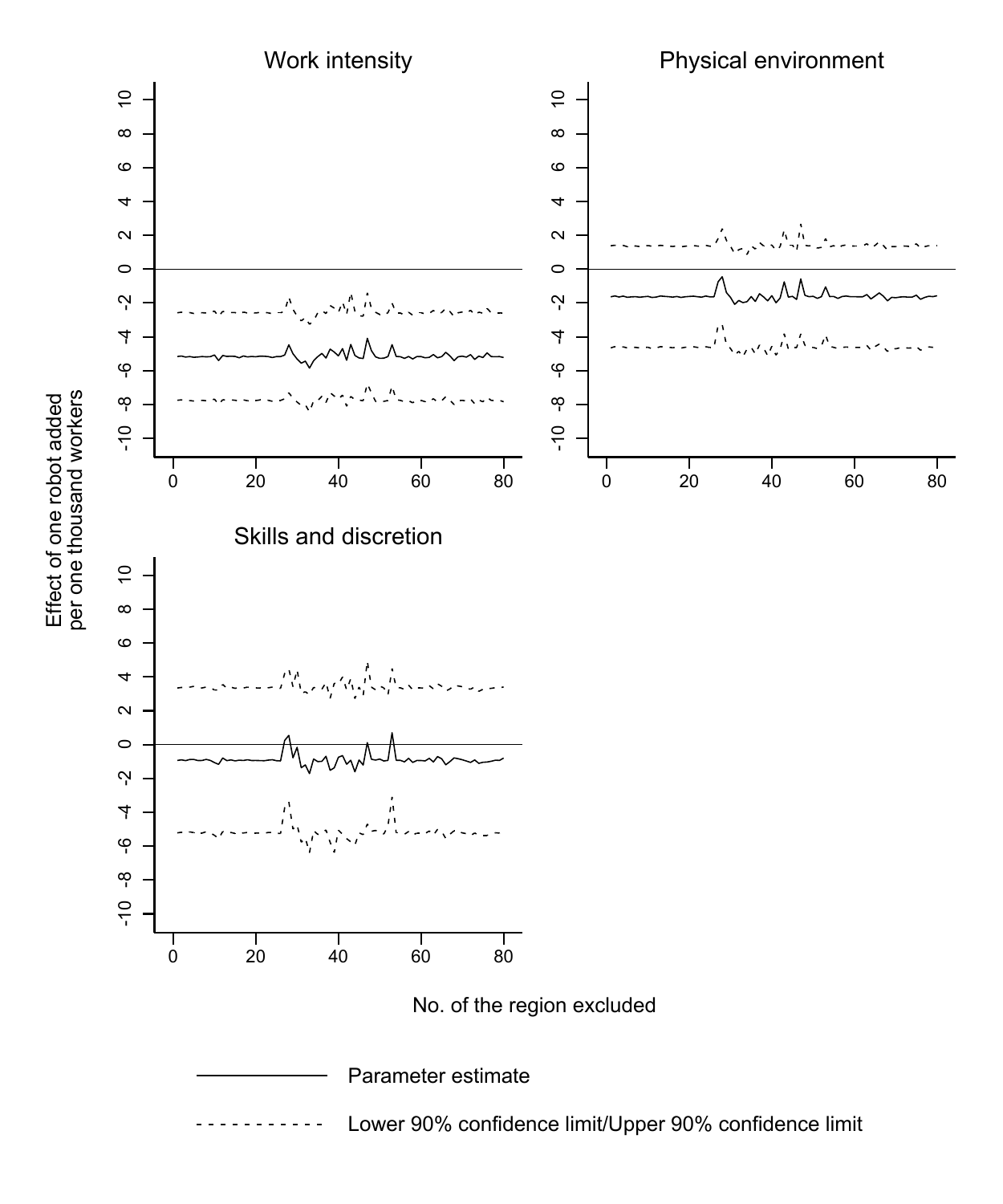} \\
		\justifying
		\footnotesize
		\noindent\textit{Note:} The figure present the parameter estimates of the effect of robotization on job quality leaving out a region each time. The econometric specifications include an intercept, a dummy for the period 2000--2005, a dummy for core-periphery European countries, start-of-period controls and controls for compositional changes. Standard errors clustered at the regional level in dashed lines. Observations weighted by the number of workers in the region at the beginning of the period.\\
		\noindent\textit{Source}: Authors' analysis from EWCS, ECHP, EU-LFS, IFR and KIP.
		\label{Figure 2}
	\end{figure}
\end{singlespace}

\clearpage

% Table 1: Descriptive statistics

\begin{singlespace}
	\begin{ThreePartTable}
		\def\sym#1{\ifmmode^{#1}\else\(^{#1}\)\fi}
		\footnotesize
		\setlength\tabcolsep{0.10em}
		\renewcommand{\arraystretch}{0.75}
		\begin{TableNotes}[flushleft]\setlength\labelsep{0pt}\footnotesize\justifying
			\item\textit{Notes}: Observations weighted by the number of workers in the region.		
			\item\textit{Source}: Authors' analysis from EWCS, ECHP, EU-LFS, IFR, EU KLEMS and WITS.
		\end{TableNotes}
		\begin{tabularx}{\textwidth}{X S[table-column-width=2.3cm]S[table-column-width=2.3cm]S[table-column-width=2.3cm]}
			\caption{Descriptive statistics} \label{Table 1}\\
			\toprule
			&\multicolumn{3}{c}{\makecell{Means \\ (standard deviations)}} \\ [2ex]	
			&\multicolumn{1}{c}{1995}&\multicolumn{1}{c}{2000}&\multicolumn{1}{c}{2005}\\
			\midrule
			Robots per thousand workers&       0.798&       1.486&       2.126\\
			&     (0.585)&     (1.192)&     (1.672)\\
			Work intensity (0--100)&      55.459&      55.374&      55.710\\
			&     (6.969)&     (6.428)&     (7.153)\\
			Quantitative demands (0--100)&      59.305&      59.878&      63.741\\
			&     (9.991)&     (7.996)&     (7.500)\\
			Pace and determinants (0--100)&      51.574&      50.859&      47.661\\
			&     (7.830)&     (9.353)&     (9.648)\\
			Physical environment (0--100)&      72.602&      70.092&      72.431\\
			&     (6.441)&     (6.889)&     (5.916)\\
			Skills and discretion (0--100)&      55.595&      53.409&      53.570\\
			&     (9.659)&     (9.290)&    (11.255)\\
			Share of workers with health affected by work&		0.618&		0.639&		0.349\\
			&		(0.149)&		(0.144)&		(0.181)\\
			Share of workers with health-related stress&		0.290&		0.228&		0.200\\
			&		(0.131)&		(0.117)&		(0.138)\\			
			Share of workers with health-related anxiety&		0.045&		0.050&		0.063\\
			&		(0.062)&		(0.062)&		(0.089)\\			
			Share of pop. employed in industry&       0.301&       0.297&       0.283\\
			&     (0.060)&     (0.067)&     (0.064)\\
			Population (thousand people)&    7061.838&    7939.785&    8014.270\\
			&  (4949.596)&  (5232.782)&  (5111.177)\\
			Share of females&       0.498&       0.498&       0.499\\
			&     (0.009)&     (0.009)&     (0.007)\\
			Share of pop. above 64&       6.654&       6.151&       5.855\\
			&     (0.941)&     (0.909)&     (0.870)\\
			Share of pop. with medium education&       0.412&       0.403&       0.420\\
			&     (0.127)&     (0.123)&     (0.120)\\
			Share of with high education&       0.168&       0.176&       0.206\\
			&     (0.065)&     (0.063)&     (0.067)\\
			Average RTI index&       0.108&       0.090&       0.031\\
			&     (0.090)&     (0.106)&     (0.071)\\
			Average offshorability index&       0.022&       0.012&      -0.052\\
			&     (0.109)&     (0.122)&     (0.098)\\
			Share of female workers&       0.215&       0.221&       0.236\\
			&     (0.117)&     (0.106)&     (0.051)\\
			Share of workers below 30&       0.224&       0.212&       0.238\\
			&     (0.123)&     (0.121)&     (0.038)\\
			Share of workers with 50 or more&       0.193&       0.236&       0.201\\
			&     (0.127)&     (0.154)&     (0.040)\\
			Share of medium educated workers&       0.404&       0.429&       0.473\\
			&     (0.178)&     (0.189)&     (0.162)\\
			Share share of highly educated workers&       0.273&       0.233&       0.173\\
			&     (0.263)&     (0.206)&     (0.078)\\
			ICT capital stock (thousand US\textdollar{} per worker)&       7.720&       6.311&       7.198\\
			&     (2.513)&     (1.556)&     (1.450)\\
			Chinese imports (US\textdollar{} per worker)&    1464.923&    3068.354&    8001.657\\
			&   (630.298)&  (1695.555)&  (4977.016) \\[1ex]
			No. of observations&\multicolumn{1}{c}{80}&\multicolumn{1}{c}{80}&\multicolumn{1}{c}{80}\\
			\bottomrule
			\insertTableNotes
		\end{tabularx}
	\end{ThreePartTable}
\end{singlespace}

\clearpage

% Table 2: Effect of robot adoption on the work intensity dimension (5-year changes, 1995--2005)

\begin{singlespace}
	\begin{ThreePartTable}
		\def\sym#1{\ifmmode^{#1}\else\(^{#1}\)\fi}
		\footnotesize
		\setlength\tabcolsep{0.05em}
		\begin{TableNotes}[flushleft]\setlength\labelsep{0pt}\footnotesize\justifying
			\item\textit{Notes}: \sym{***} significant at 1\% level; \sym{**} significant at 5\% level; \sym{*} significant at 10\% level. The left-hand-side variable in all regressions is the 5-year changes (1995--2000 and 2000--2005) in the variable of interest.
			All specifications include an intercept, a dummy for the period 2000-2005 and a dummy for core-periphery European countries. Standard errors clustered at the regional level in parentheses. Observations weighted by the number of workers in the region.		
			\item\textit{Source}: Authors' analysis from EWCS, ECHP, EU-LFS, IFR and KIP.
		\end{TableNotes}
		\begin{tabularx}{\textwidth}{X *{4}{S[table-column-width=2cm,parse-numbers=false]}}
			\caption{Effect of robot adoption on job quality: work intensity} \label{Table 2}\\
			\toprule
			&\multicolumn{1}{c}{(I)}&\multicolumn{1}{c}{(II)}&\multicolumn{1}{c}{(III)}&\multicolumn{1}{c}{(IV)}\\
			&\multicolumn{1}{c}{OLS}&\multicolumn{1}{c}{OLS}&\multicolumn{1}{c}{IV}&\multicolumn{1}{c}{IV}\\
			\midrule
			$\Delta$Robot exposure&      -4.473\sym{***}&      -4.321\sym{***}&      -5.635\sym{***}&      -5.165\sym{***}\\
			&     (1.248)         &     (1.226)         &     (1.935)         &     (1.574)         \\
			Share of employment in industry&      30.968\sym{*}  &      25.506         &      32.935\sym{*}  &      26.351         \\
			&    (17.183)         &    (18.085)         &    (17.273)         &    (17.983)         \\
			Population (log)&       2.103\sym{**} &       1.693\sym{**} &       2.068\sym{**} &       1.678\sym{**} \\
			&     (0.891)         &     (0.816)         &     (0.881)         &     (0.808)         \\
			Share of females&       3.153         &     -58.390         &      -8.109         &     -68.156         \\
			&    (68.478)         &    (88.488)         &    (70.635)         &    (91.812)         \\
			Share of pop. above 64&       0.640         &       0.657         &       0.607         &       0.613         \\
			&     (0.556)         &     (0.718)         &     (0.552)         &     (0.718)         \\
			Share of pop. with medium education&      19.641\sym{**} &      14.156\sym{*}  &      20.439\sym{***}&      14.670\sym{*}  \\
			&     (7.671)         &     (7.948)         &     (7.714)         &     (7.985)         \\
			Share of pop. with high education&      76.305\sym{***}&      56.076\sym{**} &      75.722\sym{***}&      55.436\sym{**} \\
			&    (20.201)         &    (21.979)         &    (19.954)         &    (22.377)         \\
			RTI         &      24.923\sym{**} &      10.745         &      23.343\sym{**} &       9.578         \\
			&    (10.379)         &    (12.848)         &    (10.308)         &    (13.153)         \\
			OFF         &     -20.473         &     -10.734         &     -17.765         &      -8.776         \\
			&    (13.231)         &    (14.249)         &    (13.431)         &    (14.578)         \\ [1ex]
			R\textsuperscript{2} & 0.180 & 0.267 & & \\
			No. of observations & \multicolumn{1}{c}{\hspace{6mm}\num{160}} & \multicolumn{1}{c}{\hspace{6mm}\num{160}} & \multicolumn{1}{c}{\hspace{6mm}\num{160}} & \multicolumn{1}{c}{\hspace{6mm}\num{160}} \\
			Mean of dependent variable & 0.015 & 0.015 & 0.015 & 0.015 \\
			Mean of independent variable & 0.607 & 0.607 & 0.607 & 0.607 \\
			First-stage Wald \textit{F}-statistic &&& 95.055 & 65.245 \\ [1ex]
			Compositional changes&\multicolumn{1}{c}{}&\multicolumn{1}{c}{\checkmark}&\multicolumn{1}{c}{}&\multicolumn{1}{c}{\checkmark} \\
			\bottomrule	
			\insertTableNotes
		\end{tabularx}
	\end{ThreePartTable}
\end{singlespace}

\clearpage

% Table 3: Effect of robot adoption on the physical environment and skills and discretion dimensions (5-year changes, 1995--2005)

\begin{singlespace}
	\begin{ThreePartTable}
		\def\sym#1{\ifmmode^{#1}\else\(^{#1}\)\fi}
		\footnotesize
		\setlength\tabcolsep{0.05em}
		\begin{TableNotes}[flushleft]\setlength\labelsep{0pt}\footnotesize\justifying
			\item\textit{Notes}: \sym{***} significant at 1\% level; \sym{**} significant at 5\% level; \sym{*} significant at 10\% level. The left-hand-side variable in all regressions is the 5-year changes (1995--2000 and 2000--2005) in the variable of interest.
			All specifications include an intercept, a dummy for the period 2000-2005 and a dummy for core-periphery European countries. Standard errors clustered at the regional level in parentheses. Observations weighted by the number of workers in the region.		
			\item\textit{Source}: Authors' analysis from EWCS, ECHP, EU-LFS, IFR and KIP.
		\end{TableNotes}
		\begin{tabularx}{\textwidth}{X *{4}{S[table-column-width=1.8cm]}}
			\caption{Effect of robot adoption on job quality:  physical environment and skills and discretion} \label{Table 3}\\
			\toprule
			&\multicolumn{1}{c}{(I)}&\multicolumn{1}{c}{(II)}&\multicolumn{1}{c}{(III)}&\multicolumn{1}{c}{(IV)}\\
			&\multicolumn{1}{c}{OLS}&\multicolumn{1}{c}{OLS}&\multicolumn{1}{c}{IV}&\multicolumn{1}{c}{IV}\\
			\midrule
			\textit{Panel A}. Physical environment &&&& \\[1ex]
			~~$\Delta$Robot exposure&      -0.407         &      -0.187         &      -2.081         &      -1.627         \\
			&     (1.577)         &     (1.395)         &     (2.096)         &     (1.820)         \\ [1ex]
			~~R\textsuperscript{2} & 0.142 & 0.182 & & \\
			~~No. of observations & \multicolumn{1}{c}{\num{160}} & \multicolumn{1}{c}{\num{160}} & \multicolumn{1}{c}{\num{160}} & \multicolumn{1}{c}{\num{160}} \\
			~~Mean of dependent variable & 0.463 & 0.463 & 0.463 & 0.463 \\
			~~Mean of independent variable & 0.607 & 0.607 & 0.607 & 0.607 \\
			~~First-stage Wald \textit{F}-statistic &&& 95.055 & 65.245 \\ [1ex]
			\textit{Panel B}. Skills and discretion &&&& \\[1ex]
			~~$\Delta$Robot exposure&       0.413         &       1.269         &      -1.862         &      -0.936         \\
			&     (1.863)         &     (1.811)         &     (2.883)         &     (2.606)         \\ [1ex]
			~~R\textsuperscript{2} & 0.057 & 0.067 & & \\
			~~No. of observations & \multicolumn{1}{c}{\hspace{6mm}\num{160}} & \multicolumn{1}{c}{\hspace{6mm}\num{160}} & \multicolumn{1}{c}{\hspace{6mm}\num{160}} & \multicolumn{1}{c}{\hspace{6mm}\num{160}} \\
			~~Mean of dependent variable & 0.463 & 0.463 & 0.463 & 0.463 \\
			~~Mean of independent variable & 0.607 & 0.607 & 0.607 & 0.607 \\
			~~First-stage Wald \textit{F}-statistic &&& 95.055 & 65.245 \\ [1ex]
			Start-of-period-controls&\multicolumn{1}{c}{\checkmark}&\multicolumn{1}{c}{\checkmark}&\multicolumn{1}{c}{\checkmark}&\multicolumn{1}{c}{\checkmark} \\			
			Compositional changes&\multicolumn{1}{c}{}&\multicolumn{1}{c}{\checkmark}&\multicolumn{1}{c}{}&\multicolumn{1}{c}{\checkmark} \\
			\bottomrule	
			\insertTableNotes
		\end{tabularx}
	\end{ThreePartTable}
\end{singlespace}

\clearpage

% Table 4: Effect of robot adoption on the sub-dimensions of work intensity (5-year changes, 1995--2005)

\begin{singlespace}
	\begin{table}[h!]
		\begin{ThreePartTable}
			\def\sym#1{\ifmmode^{#1}\else\(^{#1}\)\fi}
			\footnotesize
			\setlength\tabcolsep{0.10em}
			\begin{TableNotes}[flushleft]\setlength\labelsep{0pt}\footnotesize\justifying
				\item\textit{Notes}: \sym{***} significant at 1\% level; \sym{**} significant at 5\% level; \sym{*} significant at 10\% level. The left-hand-side variable in all regressions is the 5-year changes (1995--2000 and 2000--2005) in the variable of interest. All specifications include an intercept, a dummy for the period 2000-2005 and a dummy for core-periphery European countries. Standard errors clustered at the regional level in parentheses. Observations weighted by the number of workers in the region.			
				\item\textit{Source}: Authors' analysis from EWCS, ECHP, EU-LFS, IFR and KIP.
			\end{TableNotes}
			\begin{tabularx}{\linewidth}{X *{4}{S[table-column-width=2.2cm]}}
				\caption{Effect of robot adoption on the sub-dimensions of work intensity } \label{Table 4}\\
				\toprule
				&\multicolumn{2}{c}{\makecell{Quantitative\\demands}}&\multicolumn{2}{c}{\makecell{Pace and\\interdependency}}\\ [2ex]				
				&\multicolumn{1}{c}{(I)}&\multicolumn{1}{c}{(II)}&\multicolumn{1}{c}{(III)}&\multicolumn{1}{c}{(IV)}\\
				&\multicolumn{1}{c}{OLS}&\multicolumn{1}{c}{IV}&\multicolumn{1}{c}{OLS}&\multicolumn{1}{c}{IV}\\					
				\midrule
				$\Delta$Robot exposure&      -6.296\sym{***}&      -6.890\sym{***}&      -2.328\sym{**} &      -3.453\sym{*}  \\
				&     (1.670)         &     (1.938)         &     (1.161)         &     (1.780)         \\ [1ex]
				No. of observations & \multicolumn{1}{c}{\hspace{6mm}\num{160}} & \multicolumn{1}{c}{\hspace{6mm}\num{160}} & \multicolumn{1}{c}{\hspace{6mm}\num{160}} & \multicolumn{1}{c}{\hspace{6mm}\num{160}} \\
				R\textsuperscript{2} & 0.319 && 0.320\\
				Mean of dependent variable & 2.254 & 2.254 & -2.215 & -2.215\\
				Mean of independent variable & 0.607 & 0.607 & 0.607 & 0.607\\
				First-stage Wald \textit{F}-statistic && 65.245 && 65.245\\ [1ex]					
				Start-of-period controls&\multicolumn{1}{c}{\checkmark}&\multicolumn{1}{c}{\checkmark}&\multicolumn{1}{c}{\checkmark}&\multicolumn{1}{c}{\checkmark} \\ 	
				Compositional changes&\multicolumn{1}{c}{\checkmark}&\multicolumn{1}{c}{\checkmark}&\multicolumn{1}{c}{\checkmark}&\multicolumn{1}{c}{\checkmark} \\	
				\bottomrule	
				\insertTableNotes
			\end{tabularx}
		\end{ThreePartTable}
	\end{table}
\end{singlespace}

\clearpage

% Table 5: Effects on employment
\begin{mdframed}[rightline=false,leftline=false,topline=false,bottomline=false,skipabove=0pt,skipbelow=0pt]
\begin{singlespace}
	\begin{ThreePartTable}
		\def\sym#1{\ifmmode^{#1}\else\(^{#1}\)\fi}
		\footnotesize
		\setlength\tabcolsep{0.02em}
		\begin{TableNotes}[flushleft]\setlength\labelsep{0pt}\footnotesize\justifying
			\item\textit{Notes}: \sym{***} significant at 1\% level; \sym{**} significant at 5\% level; \sym{*} significant at 10\% level. The left-hand-side variable in all regressions is the 5-year changes (1995--2000 and 2000--2005) in the variable of interest. All specifications include an intercept, a dummy for the period 2000--2005, a dummy for core-periphery European countries and start-of-period controls. Standard errors clustered at the regional level in parentheses. Observations weighted by the working-age population in the region at the beginning of the period.		
			\item\textit{Source}: Authors' analysis from ECHP, EU-LFS, IFR and KIP.
		\end{TableNotes}
		\begin{tabularx}{\textwidth}{X *{6}{S[table-column-width=1.6cm,parse-numbers=false]}}
			\caption{Effect of robot adoption on employment (1995--2005)} \label{Table 5}\\
			\toprule
			&\multicolumn{2}{c}{\makecell{Employment rate\\in industry}}&\multicolumn{2}{c}{\makecell{Employment rate\\in other sectors}}&\multicolumn{2}{c}{\makecell{Overall employ-\\ment rate}}\\ [2ex]
			&\multicolumn{1}{c}{(I)}&\multicolumn{1}{c}{(II)}&\multicolumn{1}{c}{(III)}&\multicolumn{1}{c}{(IV)}&\multicolumn{1}{c}{(V)}&\multicolumn{1}{c}{(VI)}\\
			&\multicolumn{1}{c}{OLS}&\multicolumn{1}{c}{IV}&\multicolumn{1}{c}{OLS}&\multicolumn{1}{c}{IV}&\multicolumn{1}{c}{OLS}&\multicolumn{1}{c}{IV}\\
			\midrule
			$\Delta$Robot exposure&      -0.002         &      -0.007         &       0.008         &       0.011         &       0.006         &       0.004         \\
			&     (0.004)         &     (0.006)         &     (0.008)         &     (0.009)         &     (0.006)         &     (0.007)         \\ [1ex]
			R\textsuperscript{2} & 0.360 & & 0.266 & & 0.421 & \\
			No. of observations & \multicolumn{1}{c}{\hspace{6mm}\num{160}} & \multicolumn{1}{c}{\hspace{6mm}\num{160}} & \multicolumn{1}{c}{\hspace{6mm}\num{160}} & \multicolumn{1}{c}{\hspace{6mm}\num{160}} & \multicolumn{1}{c}{\hspace{6mm}\num{160}}  & \multicolumn{1}{c}{\hspace{6mm}\num{160}}\\
			Mean of dependent variable & -0.001 & -0.001 & 0.028 & 0.028 & 0.027 & 0.027\\
			Mean of independent variable & 0.557 & 0.557 & 0.557 & 0.557 & 0.557 & 0.557 \\
			\nth{1} stage Wald \textit{F} statistic &&90.593&&90.593 &&90.593\\ [1ex]			
			Start-of-period controls&\multicolumn{1}{c}{\checkmark}&\multicolumn{1}{c}{\checkmark}&\multicolumn{1}{c}{\checkmark}&\multicolumn{1}{c}{\checkmark}&\multicolumn{1}{c}{\checkmark}&\multicolumn{1}{c}{\checkmark}\\ 				
			\bottomrule	
			\insertTableNotes
		\end{tabularx}
	\end{ThreePartTable}
\end{singlespace}
\end{mdframed}

\clearpage

% Table 6: Robustness checks (work intensity, 5-year changes, 1995--2005)

\begin{landscape}
	\vspace*{\fill}	
	\begin{singlespace}
		\begin{table}[h!]
			\begin{ThreePartTable}
				\def\sym#1{\ifmmode^{#1}\else\(^{#1}\)\fi}
				\footnotesize
				\begin{TableNotes}[flushleft]\setlength\labelsep{0pt}\footnotesize\justifying
					\item\textit{Notes}: \sym{***} significant at 1\% level; \sym{**} significant at 5\% level; \sym{*} significant at 10\% level. The left-hand-side variable in all regressions is the 5-year changes (1995--2000 and 2000--2005) in the variable of interest. All specifications include an intercept, a dummy for the period 2000-2005 and a dummy for core-periphery European countries.Standard errors clustered at the regional level in parentheses. 			
					\item\textit{Source}: Authors' analysis from EWCS, ECHP, EU-LFS, IFR, KIP,  Australian labour force statistics, EUKLEMS and WITS.
				\end{TableNotes}
				\begin{tabularx}{\linewidth}{X *{5}{S[table-column-width=2.5cm]}}
					\caption{Robustness checks: work intensity (IV estimates)} \label{Table 6}\\
					\toprule
					&\multicolumn{1}{c}{\makecell{Alternative IVs\\(AUS, CH)}}&\multicolumn{1}{c}{\makecell{Alternative IV\\(SE)}}&\multicolumn{1}{c}{\makecell{Additional\\controls}}&\multicolumn{1}{c}{\makecell{Unweighted}}&\multicolumn{1}{c}{\makecell{Falsification\\test}}\\ [2ex]
					&\multicolumn{1}{c}{(I)}&\multicolumn{1}{c}{(II)}&\multicolumn{1}{c}{(III)}&\multicolumn{1}{c}{(IV)}&\multicolumn{1}{c}{(V)}\\	
					\midrule
					$\Delta$Robot exposure&      -3.903\sym{**} &      -4.631\sym{***}&      -6.567\sym{***}&      -3.387\sym{*}  &       1.067         \\
					&     (1.561)         &     (1.626)         &     (1.802)         &     (1.870)         &     (1.281)         \\ [1ex]
					No. of observations & \multicolumn{1}{c}{\hspace{6mm}\num{160}} & \multicolumn{1}{c}{\hspace{6mm}\num{158}} & \multicolumn{1}{c}{\hspace{6mm}\num{150}} & \multicolumn{1}{c}{\hspace{6mm}\num{160}} & \multicolumn{1}{c}{\hspace{6mm}\num{160}} \\
					Mean of dependent variable & 0.015 & -0.078 & 0.103 & 0.563 & 0.324 \\
					Mean of independent variable & 0.607 & 0.610 & 0.626 & 0.483 & 0.533 \\
					First-stage Wald \textit{F}-statistic & 64.607 & 50.272 & 57.702 & 70.723 & 96.571 \\
					Hansen \textit{J} \textit{p}-value & 0.401 &&&& \\ [1ex]
					Start-of-period controls&\multicolumn{1}{c}{\checkmark}&\multicolumn{1}{c}{\checkmark}&\multicolumn{1}{c}{\checkmark}&\multicolumn{1}{c}{\checkmark}&\multicolumn{1}{c}{\checkmark} \\ 	
					Compositional changes&\multicolumn{1}{c}{\checkmark}&\multicolumn{1}{c}{\checkmark}&\multicolumn{1}{c}{\checkmark}&\multicolumn{1}{c}{\checkmark}&\multicolumn{1}{c}{\checkmark} \\
					Chinese import exposure&\multicolumn{1}{c}{}&\multicolumn{1}{c}{}&\multicolumn{1}{c}{\checkmark}&\multicolumn{1}{c}{}&\multicolumn{1}{c}{} \\
					$\Delta$ICT capital&\multicolumn{1}{c}{}&\multicolumn{1}{c}{}&\multicolumn{1}{c}{\checkmark}&\multicolumn{1}{c}{}&\multicolumn{1}{c}{} \\ 					
					\bottomrule	
					\insertTableNotes
				\end{tabularx}
			\end{ThreePartTable}
		\end{table}
	\end{singlespace}
	\vspace*{\fill}	
\end{landscape}

% Table 7: Impact on health variables

\begin{singlespace}
	\begin{table}[h!]
		\begin{ThreePartTable}
			\def\sym#1{\ifmmode^{#1}\else\(^{#1}\)\fi}
			\footnotesize
			\setlength\tabcolsep{0.10em}
			\begin{TableNotes}[flushleft]\setlength\labelsep{0pt}\footnotesize\justifying
				\item\textit{Notes}: \sym{***} significant at 1\% level; \sym{**} significant at 5\% level; \sym{*} significant at 10\% level. The left-hand-side variable in all regressions is the 5-year changes (1995--2000 and 2000--2005) in the variable of interest. All specifications include an intercept, a dummy for the period 2000-2005 and a dummy for core-periphery European countries. Standard errors clustered at the regional level in parentheses. Observations weighted by the number of workers in the region.			
				\item\textit{Source}: Authors' analysis from EWCS, ECHP, EU-LFS, IFR and KIP.
			\end{TableNotes}
			\begin{tabularx}{\linewidth}{X *{3}{S[table-column-width=2.5cm]}}
				\caption{Effect of robot adoption on stress, anxiety and the share of workers reporting that work affects their health (IV estimates)} \label{Table 7}\\
				\toprule
				&\multicolumn{1}{c}{\makecell{Work affects\\health}}& \multicolumn{1}{c}{Stress}&\multicolumn{1}{c}{Anxiety}\\ 			
				&\multicolumn{1}{c}{(I)}&\multicolumn{1}{c}{(II)}&\multicolumn{1}{c}{(III)}\\					
				\midrule
				$\Delta$Robot exposure&    0.066\sym{*}&	  0.092\sym{***}&      0.032\sym{**}\\
				&		(0.034)&		(0.035)&     (0.014)\\ [1ex]
				No. of observations & \multicolumn{1}{c}{\hspace{6mm}\num{160}} & \multicolumn{1}{c}{\hspace{6mm}\num{160}} & \multicolumn{1}{c}{\hspace{6mm}\num{160}} \\
				Mean of dependent variable & -0.028 & -0.040 & 0.004 \\
				Mean of independent variable & 0.607 & 0.607 & 0.607\\
				First-stage Wald \textit{F}-statistic & 65.245 & 65.245 & 65.245\\ [1ex]					
				Start-of-period controls&\multicolumn{1}{c}{\checkmark}&\multicolumn{1}{c}{\checkmark}&\multicolumn{1}{c}{\checkmark}\\ 	
				Compositional changes&\multicolumn{1}{c}{\checkmark}&\multicolumn{1}{c}{\checkmark}&\multicolumn{1}{c}{\checkmark}\\	
				\bottomrule	
				\insertTableNotes
			\end{tabularx}
		\end{ThreePartTable}
	\end{table}
\end{singlespace}
\clearpage

\appendix
\section*{Annex}
\setcounter{table}{0}
\renewcommand\thetable{A\arabic{table}}

% Table A1: First-stage regression (5-year changes, 1995--2005)

\begin{singlespace}
	\begin{ThreePartTable}
		\def\sym#1{\ifmmode^{#1}\else\(^{#1}\)\fi}
		\footnotesize
		\begin{TableNotes}[flushleft]\setlength\labelsep{0pt}\footnotesize\justifying
			\item\textit{Notes}: \sym{***} significant at 1\% level; \sym{**} significant at 5\% level; \sym{*} significant at 10\% level. The left-hand-side variable in all regressions is the 5-year changes (1995--2000 and 2000--2005) in the variable of interest.
			All specifications include an intercept, a dummy for the period 2000-2005 and a dummy for core-periphery European countries. Standard errors clustered at the regional level in parentheses. Observations weighted by the number of workers in the region.
			\item\textit{Source}: Authors' analysis from EWCS, ECHP, EU-LFS, IFR and KIP.
		\end{TableNotes}
		\begin{tabularx}{\textwidth}{X *{2}{S[table-column-width=3cm]}}
			\caption{First-stage regression of the change in robot exposure on the change in robot exposure using the changes in sectoral robot density in South Korea} \label{Table A1}\\
			\toprule
			&\multicolumn{1}{c}{(I)}&\multicolumn{1}{c}{(II)}\\
			\midrule
			$\Delta$Robot exposure (South Korea)&       1.057\sym{***}&       1.073\sym{***}\\
			&     (0.104)         &     (0.125)         \\
			Share of industry&      -0.540         &      -0.939         \\
			&     (0.744)         &     (0.819)         \\
			Population (log)&      -0.028         &      -0.031         \\
			&     (0.026)         &     (0.029)         \\
			Share of females&       2.606         &       2.701         \\
			&     (2.585)         &     (2.691)         \\
			Share of pop. above 64&      -0.052\sym{***}&      -0.052\sym{***}\\
			&     (0.018)         &     (0.017)         \\
			Share of pop. with medium education&       1.445\sym{***}&       1.547\sym{***}\\
			&     (0.310)         &     (0.332)         \\
			Share of pop. with high education&       0.186         &      -0.462         \\
			&     (0.468)         &     (0.605)         \\
			RTI         &      -1.265\sym{***}&      -1.666\sym{***}\\
			&     (0.295)         &     (0.346)         \\
			OFF         &       0.779\sym{**} &       1.120\sym{***}\\
			&     (0.335)         &     (0.392)         \\ [1ex]
			R\textsuperscript{2} & 0.849 & 0.856 \\
			No. of observations & \multicolumn{1}{c}{\hspace{6mm}\num{160}} & \multicolumn{1}{c}{\hspace{6mm}\num{160}} \\
			Mean of dependent variable & 0.607 & 0.607 \\
			Mean of the instrument & 0.793 & 0.793 \\
			First-stage Wald \textit{F}-statistic & 95.055 & 65.245 \\
			Partial R\textsuperscript{2} of instrument & 0.727 & 0.708 \\ [1ex]		
			Compositional changes&\multicolumn{1}{c}{}&\multicolumn{1}{c}{\checkmark} \\
			\bottomrule	
			\insertTableNotes
		\end{tabularx}
	\end{ThreePartTable}
\end{singlespace}

\clearpage

% Table A2: Results for 2005--2015

\begin{singlespace}
	\begin{ThreePartTable}
		\def\sym#1{\ifmmode^{#1}\else\(^{#1}\)\fi}
		\footnotesize
		\setlength\tabcolsep{0.05em}
		\begin{TableNotes}[flushleft]\setlength\labelsep{0pt}\footnotesize\justifying
			\item\textit{Notes}: \sym{***} significant at 1\% level; \sym{**} significant at 5\% level; \sym{*} significant at 10\% level. The left-hand-side variable in all regressions is the 5-year changes (2005--2010 and 2010--2015) in the variable of interest.
			All specifications include an intercept, a dummy for the period 2010--2015 and a dummy for core-periphery European countries. Standard errors clustered at the regional level in parentheses. Observations weighted by the number of workers in the region at the beginning of the period.		
			\item\textit{Source}: Authors' analysis from EWCS, ECHP, EU-LFS, IFR and KIP.
		\end{TableNotes}
		\begin{tabularx}{\textwidth}{X *{4}{S[table-column-width=2cm,parse-numbers=false]}}
			\caption{Effect of robot adoption on job quality (2005--2015)} \label{Table A2}\\
			\toprule
			&\multicolumn{1}{c}{(I)}&\multicolumn{1}{c}{(II)}&\multicolumn{1}{c}{(III)}&\multicolumn{1}{c}{(IV)}\\
			&\multicolumn{1}{c}{OLS}&\multicolumn{1}{c}{OLS}&\multicolumn{1}{c}{IV}&\multicolumn{1}{c}{IV}\\
			\midrule
			\textit{Panel A}. Job intensity &&&&\\[1ex]
			~~$\Delta$Robot exposure&       2.467         &       2.494         &       7.064         &       7.109         \\
			&     (2.427)         &     (2.634)         &     (7.004)         &     (6.529)         \\[1ex]	
			~~R\textsuperscript{2} & 0.062 & 0.094 & & \\
			~~No. of observations & \multicolumn{1}{c}{\hspace{6mm}\num{160}} & \multicolumn{1}{c}{\hspace{6mm}\num{160}} & \multicolumn{1}{c}{\hspace{6mm}\num{160}} & \multicolumn{1}{c}{\hspace{6mm}\num{160}} \\
			~~Mean of dependent variable & 1.626 & 1.626 & 1.626 & 1.626 \\
			~~Mean of independent variable & 0.334 & 0.334 & 0.334 & 0.334 \\
			~~First-stage Wald \textit{F} statistic &&& 11.628 & 14.814 \\[2ex]
			\textit{Panel B}. Physical environment &&&&\\[1ex]
			~~$\Delta$Robot exposure&       0.290         &       0.292         &       2.869         &       1.079         \\
			&     (2.199)         &     (2.051)         &     (7.162)         &     (6.185)         \\[1ex]
			~~R\textsuperscript{2} & 0.019 & 0.117 & & \\
			~~No. of observations & \multicolumn{1}{c}{\hspace{6mm}\num{160}} & \multicolumn{1}{c}{\hspace{6mm}\num{160}} & \multicolumn{1}{c}{\hspace{6mm}\num{160}} & \multicolumn{1}{c}{\hspace{6mm}\num{160}} \\
			~~Mean of dependent variable & 0.386 & 0.386 & 0.386 & 0.386 \\
			~~Mean of independent variable & 0.334 & 0.334 & 0.334 & 0.334 \\
			~~First-stage Wald \textit{F} statistic &&& 11.628 & 14.814 \\[2ex]
			\textit{Panel C}. Skills and discretion &&&&\\[1ex]
			$\Delta$Robot exposure&      -3.424         &      -3.288         &      -5.828         &      -4.801         \\
			&     (2.258)         &     (2.285)         &     (4.640)         &     (4.031)         \\[1ex]
			~~R\textsuperscript{2} & 0.081 & 0.110 & & \\
			~~No. of observations & \multicolumn{1}{c}{\hspace{6mm}\num{160}} & \multicolumn{1}{c}{\hspace{6mm}\num{160}} & \multicolumn{1}{c}{\hspace{6mm}\num{160}} & \multicolumn{1}{c}{\hspace{6mm}\num{160}} \\
			~~Mean of dependent variable & 3.832 & 3.832 & 3.832 & 3.832 \\
			~~Mean of independent variable & 0.334 & 0.334 & 0.334 & 0.334 \\
			~~First-stage Wald \textit{F} statistic &&& 11.628 & 14.814 \\[1ex]
			Start-of-period controls&\multicolumn{1}{c}{\checkmark}&\multicolumn{1}{c}{\checkmark}&\multicolumn{1}{c}{\checkmark}&\multicolumn{1}{c}{\checkmark} \\
			Compositional changes&\multicolumn{1}{c}{}&\multicolumn{1}{c}{\checkmark}&\multicolumn{1}{c}{}&\multicolumn{1}{c}{\checkmark} \\
			\bottomrule	
			\insertTableNotes
		\end{tabularx}
	\end{ThreePartTable}
\end{singlespace}

\clearpage

% Table A3: Robustness checks (5-year changes, 1995--2005)

\begin{landscape}
	\vspace*{\fill}	
	\begin{singlespace}
		\begin{table}[h!]
			\begin{ThreePartTable}
				\def\sym#1{\ifmmode^{#1}\else\(^{#1}\)\fi}
				\footnotesize
				\begin{TableNotes}[flushleft]\setlength\labelsep{0pt}\footnotesize\justifying
					\item\textit{Notes}: \sym{***} significant at 1\% level; \sym{**} significant at 5\% level; \sym{*} significant at 10\% level. The left-hand-side variable in all regressions is the 5-year changes (1995--2000 and 2000--2005) in the variable of interest. All specifications include an intercept, a dummy for the period 2000-2005 and a dummy for core-periphery European countries.Standard errors clustered at the regional level in parentheses. 		
					\item\textit{Source}: Authors' analysis from EWCS, ECHP, EU-LFS, IFR, KIP,  Australian labour force statistics, EUKLEMS and WITS.
				\end{TableNotes}
				\begin{tabularx}{\linewidth}{X *{5}{S[table-column-width=2.5cm]}}
					\caption{Robustness checks: physical environment (IV estimates)} \label{Table A3}\\
					\toprule
					&\multicolumn{1}{c}{\makecell{Alternative IVs\\(AUS, CH)}}&\multicolumn{1}{c}{\makecell{Alternative IV\\(SE)}}&\multicolumn{1}{c}{\makecell{Additional\\controls}}&\multicolumn{1}{c}{\makecell{Unweighted}}&\multicolumn{1}{c}{\makecell{Falsification\\test}}\\ [2ex]
					&\multicolumn{1}{c}{(I)}&\multicolumn{1}{c}{(II)}&\multicolumn{1}{c}{(III)}&\multicolumn{1}{c}{(IV)}&\multicolumn{1}{c}{(V)}\\
					\midrule
					$\Delta$Robot exposure&      -1.052         &      -0.810         &      -0.589         &      -0.591         &      -0.810         \\
					&     (0.803)         &     (0.855)         &     (0.946)         &     (1.086)         &     (0.842)         \\[1ex]
					No. of observations & \multicolumn{1}{c}{\hspace{6mm}\num{160}} & \multicolumn{1}{c}{\hspace{6mm}\num{158}} & \multicolumn{1}{c}{\hspace{6mm}\num{150}} & \multicolumn{1}{c}{\hspace{6mm}\num{160}} & \multicolumn{1}{c}{\hspace{6mm}\num{160}} \\
					Mean of dependent variable & 1.129 & 1.180 & 1.150 & 1.020 & 1.129 \\
					Mean of independent variable & 0.533 & 0.535 & 0.545 & 0.533 & 0.533 \\
					First-stage Wald \textit{F}-statistic & 70.187 & 84.057 & 85.491 & 88.115 & 96.571 \\
					Hansen \textit{J} \textit{p}-value & 0.943 &&&& \\[1ex]	
					Start-of-period controls&\multicolumn{1}{c}{\checkmark}&\multicolumn{1}{c}{\checkmark}&\multicolumn{1}{c}{\checkmark}&\multicolumn{1}{c}{\checkmark}&\multicolumn{1}{c}{\checkmark} \\ 	
					Compositional changes&\multicolumn{1}{c}{\checkmark}&\multicolumn{1}{c}{\checkmark}&\multicolumn{1}{c}{\checkmark}&\multicolumn{1}{c}{\checkmark}&\multicolumn{1}{c}{\checkmark} \\
					Chinese import exposure&\multicolumn{1}{c}{}&\multicolumn{1}{c}{}&\multicolumn{1}{c}{\checkmark}&\multicolumn{1}{c}{}&\multicolumn{1}{c}{} \\
					$\Delta$ICT capital&\multicolumn{1}{c}{}&\multicolumn{1}{c}{}&\multicolumn{1}{c}{\checkmark}&\multicolumn{1}{c}{}&\multicolumn{1}{c}{} \\ 	
					\bottomrule	
					\insertTableNotes
				\end{tabularx}
			\end{ThreePartTable}
		\end{table}
	\end{singlespace}
	\vspace*{\fill}	
\end{landscape}

% Table A4: Robustness checks (5-year changes, 1995--2005)

\begin{landscape}
	\vspace*{\fill}	
	\begin{singlespace}
		\begin{table}[h!]
			\begin{ThreePartTable}
				\def\sym#1{\ifmmode^{#1}\else\(^{#1}\)\fi}
				\footnotesize
				\begin{TableNotes}[flushleft]\setlength\labelsep{0pt}\footnotesize\justifying
					\item\textit{Notes}: \sym{***} significant at 1\% level; \sym{**} significant at 5\% level; \sym{*} significant at 10\% level. The left-hand-side variable in all regressions is the 5-year changes (1995--2000 and 2000--2005) in the variable of interest. All specifications include an intercept, a dummy for the period 2000-2005 and a dummy for core-periphery European countries. Standard errors clustered at the regional level in parentheses. 		
					\item\textit{Source}: Authors' analysis from EWCS, ECHP, EU-LFS, IFR, KIP,  Australian labour force statistics, EUKLEMS and WITS.
				\end{TableNotes}
				\begin{tabularx}{\linewidth}{X *{5}{S[table-column-width=2.5cm]}}
					\caption{Robustness checks: skills and discretion (IV estimates)} \label{Table A4}\\
					\toprule
					&\multicolumn{1}{c}{\makecell{Alternative IVs\\(AUS, CH)}}&\multicolumn{1}{c}{\makecell{Alternative IV\\(SE)}}&\multicolumn{1}{c}{\makecell{Additional\\controls}}&\multicolumn{1}{c}{\makecell{Unweighted}}&\multicolumn{1}{c}{\makecell{Falsification\\test}}\\ [2ex]
					&\multicolumn{1}{c}{(I)}&\multicolumn{1}{c}{(II)}&\multicolumn{1}{c}{(III)}&\multicolumn{1}{c}{(IV)}&\multicolumn{1}{c}{(V)}\\
					\midrule
					$\Delta$Robot exposure&       0.347         &      -0.282         &       1.063         &       2.159         &      -0.100         \\
					&     (1.408)         &     (1.472)         &     (1.536)         &     (1.992)         &     (1.496)         \\ [1ex]
					No. of observations & \multicolumn{1}{c}{\hspace{6mm}\num{160}} & \multicolumn{1}{c}{\hspace{6mm}\num{158}} & \multicolumn{1}{c}{\hspace{6mm}\num{150}} & \multicolumn{1}{c}{\hspace{6mm}\num{160}} & \multicolumn{1}{c}{\hspace{6mm}\num{160}} \\
					Mean of dependent variable & -0.140 & -0.247 & -0.211 & 0.840 & -0.140 \\
					Mean of independent variable & 0.533 & 0.535 & 0.545 & 0.533 & 0.533 \\
					First-stage Wald \textit{F}-statistic & 70.187 & 84.057 & 85.491 & 88.115 & 96.571 \\
					Hansen \textit{J} \textit{p}-value & 0.868 &&&& \\ [1ex]					
					Start-of-period controls&\multicolumn{1}{c}{\checkmark}&\multicolumn{1}{c}{\checkmark}&\multicolumn{1}{c}{\checkmark}&\multicolumn{1}{c}{\checkmark}&\multicolumn{1}{c}{\checkmark} \\ 	
					Compositional changes&\multicolumn{1}{c}{\checkmark}&\multicolumn{1}{c}{\checkmark}&\multicolumn{1}{c}{\checkmark}&\multicolumn{1}{c}{\checkmark}&\multicolumn{1}{c}{\checkmark} \\
					Chinese import exposure&\multicolumn{1}{c}{}&\multicolumn{1}{c}{}&\multicolumn{1}{c}{\checkmark}&\multicolumn{1}{c}{}&\multicolumn{1}{c}{} \\
					$\Delta$ICT capital&\multicolumn{1}{c}{}&\multicolumn{1}{c}{}&\multicolumn{1}{c}{\checkmark}&\multicolumn{1}{c}{}&\multicolumn{1}{c}{} \\ 						
					\bottomrule	
					\insertTableNotes
				\end{tabularx}
			\end{ThreePartTable}
		\end{table}
	\end{singlespace}
	\vspace*{\fill}	
\end{landscape}

\end{document}